\documentclass[twocolumn,showpacs,preprintnumbers,superscriptaddress,amsmath,amssymb]{revtex4}
\usepackage{graphicx}
\usepackage{dcolumn}
\usepackage{bm}

\begin{document}

\title{\bf Plasma Suppression of Large Scale Structure Formation in the Universe} 

\author{Pisin Chen}%
 \email{chen@slac.stanford.edu}
\affiliation{Kavli Institute for Particle Astrophysics and
Cosmology,\\
Stanford Linear Accelerator Center, Stanford University, Stanford, CA 94309}%

\author{Kwang-Chang Lai}
\affiliation{INAF Osservatorio Astrofisico di Arcetri, Largo E.
Fermi 5, Firenze, 50125, Italy.}

\date{\today}

\begin{abstract}
We point out that during the reionization epoch of the cosmic
history, the plasma collective effect among the ordinary matter
would suppress the large scale structure formation. The imperfect
Debye shielding at finite temperature would induce a residual
long-range electrostatic potential which, working together with
the baryon thermal pressure, would counter the gravitational
collapse. As a result the effective Jean's length,
$\tilde{\lambda}_J$, is increased by a factor,
$\tilde{\lambda}_J/\lambda_J=\sqrt{8/5}$, relative to the conventional
one. For scales smaller than the effective Jean's scale the plasma
would oscillate at the ion-acoustic frequency. The modes that
would be influenced by this effect depend on the starting time and the initial temperature of reionization, but roughly lie in the range
$0.5 h{\rm Mpc}^{-1}< k$, which corresponds to the region of the
Lyman-$\alpha$ forest from the inter-galactic medium. We predict
that in the linear regime of density-contrast growth, the plasma
suppression of the matter power spectrum would approach
$1-(\Omega_{dm}/\Omega_m)^2\sim 1-(5/6)^2\sim 30\%$.
\end{abstract}



\maketitle

The matter power spectrum of the large scale structure, alongside
with the cosmic microwave background (CMB) fluctuations, provides
important information on the composition and the evolution history
of the universe. Unlike the CMB fluctuations, which was frozen
since the recombination time, the matter power spectrum continued to
evolve until now, during which the universe has
been essentially in a reionization, or plasma, state. Plasmas are
known to support numerous electron and ion collective oscillation
modes. Although in the pre-recombination era the universe was also
in a plasma state, the extreme dominance of radiation pressure
over that of matter (by a factor $10^{10}$) before decoupling renders the
contribution of ($ep$) plasma oscillations totally negligible. In
the post-decoupling era, the CMB pressure became
negligible. Thus during the reionization epoch\cite{Barkana:2006}
one might expect collective plasma effects to exhibit influence on
the structure formation.

A perfectly uniform plasma is charge-neutral and has no long-range
electrostatic potential. The inevitable thermal fluctuations of
its electron density at finite temperature would, however, induce
a non-vanishing residual electrostatic potential among the ions that
defies the Debye shielding. Such a residual electric force that exerts
an additional pressure on the ions is non-collisional in
nature. When combined with the collision-induced thermal pressure
among them, these pressures can support ion-acoustic oscillations,
a phenomenon well-known in plasma
physics.\cite{plasma:chen,plasma:krall}

In this Letter we point out that the inclusion of such plasma
collective effects would result in an increase of the sound speed
and therefore an increase of the Jean's scale by a factor
$\tilde{\lambda}_J/\lambda_J= \sqrt{8/5}$. The modes that would be
influenced by this effect depend on the starting time of reioniation
and its initial temperature, but should roughly lie in the range $0.5 h{\rm
Mpc}^{-1}< k$, which corresponds to the region of the
Lyman-$\alpha$ forest. This plasma suppression mechanism is effective to
both linear and nonlinear regimes of density perturbations. We
predict that, the maximum suppression of the power spectrum in the
linear regime should approach $\sim 1-(\Omega_{dm}/\Omega_m)^2\sim
1-(5/6)^2\sim 30\%$.

When the electrostatic potential is included, the standard
Einstein-Boltzmann equation in the Newtonian limit is now extended
into a Maxwell-Einstein-Boltzmann (MEB) equation. Let us designate
the baryon perturbation as $\Delta\equiv \delta\rho_b/\rho_b$. In
the electrostatic and the Newtonian limits and with the density
perturbation $\Delta\ll 1$, the MEB equation reads,
\begin{equation}
\ddot{\Delta}+2H\dot{\Delta}-\nabla^2\Big[\frac{\delta P}{\rho_b}
+\frac{e\phi_{em}}{m_b}\Big]=4\pi G\rho\Delta\,
.\label{perturb-evolution}
\end{equation}
where the electrostatic potential satisfies the Poisson equation:
$ \nabla^2\phi_{em}=4\pi e\rho_e$. The source of gravity contains
both dark matter and baryons, i.e., $\rho=\rho_{dm}+\rho_{b}$. We
assume that the reionization epoch began at a certain time $a^*$.
The adiabatic condition dictates that the initial perturbations
are identical between the dark and the baryonic matters, i.e.,
$\delta\rho_{dm}(a^*)/\rho_{dm}(a^*)=\delta\rho_{b}(a^*)/\rho_{b}(a^*)
=\delta\rho(a^*)/\rho(a^*)\equiv\Delta^*$. After that, dark matter and baryonic matter would still evolve
hand-in-hand and grow linearly for scales larger than the Jean's
scale, but would evolve separately for scales smaller than that. It is this latter situation which we are
focusing for our plasma effect. Thermodynamics governs that
\begin{equation}
\frac{\delta P}{\rho_b}=\frac{\delta
P}{\delta\rho_b}\frac{\delta\rho_b}{\rho_b}=\frac{\gamma_bk_BT_b}{m_b}\Delta\,
.
\end{equation}
On the other hand, the electrostatic potential at a given
temperature is given by
\begin{equation}
e\phi_{\rm em}=\frac{4\pi
e^2\delta\rho_e}{k^2+\lambda_D^{-2}}\approx \gamma_ek_BT_e\Delta\,
.
\end{equation}
The approximation results from the fact that the typical cosmic
scale of interest is much larger than the Debye length, i.e.,
$\lambda\sim 1/k \gg \lambda_D\equiv \sqrt{\gamma_ek_BT_e/4\pi
e^2\rho_e}$. We have further assumed that
$\delta\rho_e/\rho_e\simeq\delta\rho_b/\rho_b=\Delta$. This is
because globally the fast-moving electrons tend to rapidly
readjust themselves to follow the baryon density perturbations. At
first sight, it may seem counter-intuitive that the charge
neutrality condition would allow for a non-vanishing electrostatic
potential. In reality, due to the inevitable thermal fluctuations
of the much lighter electrons the plasma is never perfectly
neutral. Balance of forces on electrons requires that the electron
number density satisfies the Boltzmann's relation:
$n_e=\rho_e/m_e=n_{e0}\exp(-e\phi_{em}/k_BT_e)$. Thus at finite
temperature the Debye shielding is never perfect, and there is
always a residual electrostatic potential at the level of
$e\phi_{em}\sim k_BT_e\delta\rho_e/\rho_e$ in the leading order of
Taylor expansion. This, that one assumes
$\delta\rho_e/\rho_e=\delta\rho_b/\rho_b$ and $\nabla\cdot
E=\nabla^2\phi_{em}\neq 0$ at the same time, is the so-called
quasineutrality, or {\it plasma approximation} in plasma
physics\cite{plasma:chen}.

Fourier transforming Eq.(\ref{perturb-evolution}) leads to
\begin{equation}
\ddot{\Delta}_k+2H\dot{\Delta}_k+\Big(\frac{k}{a}\Big)^2\tilde{c}^2_{s}\Delta_k=4\pi
G\rho\Delta_k\,\label{MEB},
\end{equation}
where
\begin{equation}
\tilde{c}_s\equiv \Big(\frac{\gamma_ek_BT_e+\gamma_b k_B
T_b}{m_b}\Big)^{1/2}=\frac{\omega}{k} \label{sound dispersion}
\end{equation}
is the `sound speed' of the ion-acoustic waves. We see that the main change from the ordinary acoustic wave to the
ion acoustic wave is that there is an extra contribution to the
sound speed from the electron temperature. The fast-moving
electrons are isothermal. So $\gamma_e=1$. But for protons,
$\gamma_b=5/3$. In the case where electrons and ions are in
thermal equilibrium, i.e., $T_e=T_b$, the net change is an
increase of the sound speed from the ordinary acoustic wave,
$\sqrt{5/3}$, to that of the ion-acoustic wave, $\sqrt{1+5/3}$.
Both waves, however, are constant-velocity pressure waves that
satisfy linear dispersion relations in the long wavelength limit.

So far the temperature and the on-set of the reionization are
still unresolved. In the literature (e.g.
\cite{Benson,Tegmark,Iliev}), the reionization temperature ranges
from $10^3{\rm K}$ to $10^7{\rm K}$. As for the on-set of
reionization, the observation of the Gunn-Peterson trough in the
spectra of high redshift quasars indicates that the reionization
process was completed by $z^*\approx 6$ \cite{Becker,Fan}, while
WMAP determines that $z^*=17\pm5$ \cite{Kogut,Spergel}. For
simplicity and without compromising qualitative features of our
effect, we model the reionization as an abrupt transition at a
given instant $a^*=1/(1+z^*)$ where the plasma temperature raised
instantly to $T_b^*$ homogeneously throughout the universe.

To solve Eq.(\ref{MEB}) for a given $k$-mode, it is essential
to recognize its relationship with the initial and the final Jean's scales at
$a^*$ and $a$, respectively. Since the universe
expands adiabatically during the matter-dominant epoch, the
large-scale baryon/plasma temperature satisfies the equation of
state, $T_bV^{\gamma_b-1}={\rm const.}$, where $\gamma_b=5/3$ and
$V\propto 1/\rho \propto a^3$. Therefore $T_b\propto
\tilde{c}_s^2\propto a^{-2}$. Thus the post-reionization ($a
> a^*$) Jean's scale evolves as
\begin{equation}
\tilde{k}_{J}(a)=\frac{\sqrt{4\pi G\rho(a)}}{\tilde{c}_s(a)/a}
                     =\Big({\frac{3\pi }{2}}\frac{m_bG\rho^*}{T_b^*}\Big)^{1/2}\sqrt{a^*a}\geq
\tilde{k}_J^*\, .
\end{equation}
There are three regimes of $k$-modes that concern us. For
$k<\tilde{k}_J^*$, the thermal and the plasma pressures are
negligible from the outset. So the baryon perturbation grows
linearly in the same way as that of the dark matter:
$\delta_{b}\propto\delta_{dm}\propto a$. The decay mode,
$\delta_{b(dm)}\propto a^{-3/2}$, fades away in time and can be
ignored.

For $k>\tilde{k}_J^*$, care must be taken regarding $k$'s further
relationship with the Jean's scale at the time of interest. If
$k>\tilde{k}_J(a)>\tilde{k}_J^*$, then the combined pressure
dominates over the gravity throughout the period $(a^*,a)$. Thus
the right-hand-side of Eq.(\ref{MEB}) is negligible. Changing
variable from $t$ to $a$, with $a \propto t^{2/3}$ in the
matter-dominant era, we find
\begin{equation}
\frac{d^2\Delta_k}{da^2}+\frac{3}{2a}\frac{d\Delta_k}{da}+\Big(\frac{k^2\tilde{c}^2_s}{aH_0^2}\Big)\Delta_k=0\,\label{MEBa}.
\end{equation}
Let us introduce a parameter $\omega_{r}$ which satisfies the
relation, $k^2\tilde{c}_s^2/H_0^2\equiv \omega_{r}^2a^{-2}$, such
that $\omega_{r}$ is independent of $a$. Since $\tilde{k}_J\propto
a^{1/2}$, we can reexpress it as
\begin{equation}
\omega_{r}^2=\frac{\tilde{c}_s^2k^2a^2}{H_0^2}\simeq
\frac{3}{2}\frac{\Omega_m}{\Omega_0}\Big(\frac{k}{\tilde{k}_J^*}\Big)^2a^*\,
.
\end{equation}
It can be shown, through one more change of variable to
$x=a^{-2}$, that the solution to Eq.(\ref{MEBa}) is oscillatory:
$\Delta_k\propto \exp(i2\omega_{r}/\sqrt{a})$. Physically
$\omega_{r}/\sqrt{a}$ represents the phase of the ripples of the
density contrast in $k$-space due to the ion-acoustic oscillation.
Such an oscillation begins at the time of reionization, $a^*$.
Although in reality $\tilde{c}_sk$ is the physical ion-acoustic
oscillation frequency, $\omega_r$ can be viewed as an effective
ion-acoustic frequency since it relates to $\tilde{c}_sk$
straight-forwardly.

Ion-acoustic oscillations are known to suffer Landau
damping\cite{plasma:chen}, which was not included in
Eq.(\ref{MEBa}) {\it a priori}. To include this effect, we invoke
a simple empirical formula\cite{plasma:chen} for the Landau
damping dispersion relation that relates the real and the
imaginary parts of the ion-acoustic oscillation frequency,
\begin{equation}
\frac{\omega_{i}}{\omega_{r}}\simeq
1.1\Big(\frac{T_e}{T_b}\Big)^{7/4}e^{-(T_e/T_b)^2}\, , \quad 1\leq
T_e/T_b \leq 10\, . \label{empirical plasma dispersion}
\end{equation}
Since the electron and proton temperatures in the reionization
epoch are roughly equal, we find that $\omega_{i}/\omega_{r}\sim
0.4$ in our case. This means that within $\sim 0.4$ of an
oscillation, the ion-acoustic wave would be Landau-damped to one
e-folding of its initial amplitude. Now we replace the
$\omega_{r}$ in the exponent of the solution to Eq.(\ref{MEBa}) by
the complex frequency which satisfies the above dispersion
relation. Imposing the initial condition at $a^*$, we arrive at
the perturbation for the $k$-mode whose scale has been below the
Jean's scale all the time from $a^*$ to $a$, i.e., $k>
\tilde{k}_J(a)>\tilde{k}_J^*$,
\begin{equation}
\Delta_k\simeq \Delta_k^*e^{(-0.8+i2)\omega_{r}\eta}\,
,\label{klarge}
\end{equation}
where $\eta=\eta(a^*,a)\equiv \sqrt{1/a^*}-\sqrt{1/a}$.

For the intermediate regime $\tilde{k}_J^*< k < \tilde{k}_J$, the
$k$-mode must have exited the Jean's scale at an intermediate time
$a_c$: $k=\tilde{k}_J(a_c)$, where $a^* < a_c < a$. Such a mode
would oscillate for a period $\Delta a=a_c-a^*$. Then it would
stop the oscillation at $a=a_c$ and resume its gravitational
collapse during the subsequent interval $a-a_c$. Nevertheless,
such a resumed growth of perturbation is delayed, because of the
ion-acoustic oscillations during the earlier time interval, by the
same amount
\begin{equation}
\Delta a\equiv
a_c-a^*=\Big[\Big(\frac{k}{\tilde{k}_J^*}\Big)^{2}-1\Big]a^*\,.
\end{equation}
Thus the evolution of such a mode $\tilde{k}_J^*< k
<\tilde{k}_J(a)$ should become
\begin{equation}
\Delta_k=\Delta_k^*\Big(\frac{a-\Delta a}{a^*}\Big)
           =\Delta_k^*\Big[\frac{a}{a^*}+1-\Big(\frac{k}{\tilde{k}_J^*}\Big)^{2}\Big]\,
           .\label{ksmall}
\end{equation}
Here we have ignored the fact that upon crossing the Jean's scale
the $k$-mode may still carry a non-vanishing oscillation
amplitude. We expect that such remaining amplitude should be small
due to Landau damping.

Our aim is to identify the imprints of this plasma effect in the
large scale structure formation through observations. So we look
for its modification to the matter power spectrum\cite{Dodelson},
\begin{equation}
P(k,a)=2\pi^2\delta_H^2\frac{k^n}{H_0^{n+3}}T^2(k)D^2(a)\,
.\label{MPS}
\end{equation}
Here $\delta_H=1.9\times10^{-5}$ is the amplitude of primordial
perturbations, $H_0=100h {\rm km}{\rm s}^{-1}{\rm Mpc}^{-1}$ the
Hubble parameter at present, $T(k)$ the transfer function, and
$D(a)$ the growth function. We limit our investigation in the
linear regime of power spectrum growth. Due to the onset of the
plasma effect, the conventional growth function, $D(a)$, in
Eq.(\ref{MPS}) has to be modified. Let us denote the modified
growth function as $\tilde{D}(k,a)$. Clearly, the plasma
suppression would only affect the regime where $k> \tilde{k}_J^*$,
which is what we will concentrate below. To track the plasma
effect on the baryons for $a>a^*$, we write
\begin{eqnarray}
\tilde{D}(k,a)&=&D_{dm}(a)\frac{\Omega_{dm}}{\Omega_m}+D_b(a)\frac{\Omega_b}{\Omega_m}\nonumber
\\
&=&D_{dm}(a^*)\frac{a}{a^*}\frac{\Omega_{dm}}{\Omega_m}+D_b(a^*)\frac{\tilde{\Delta}_k}{\Delta_k^*}\frac{\Omega_b}{\Omega_m}
\, .
\end{eqnarray}
Invoking the initial condition, $D_{dm}(a^*)= D_b(a^*)=D(a^*)$,
and under the conventional notion of linear growth,
$D(a)=D(a^*)(a/a^*)$, we arrive at the relative change of the
growth function as
\begin{equation}
\frac{\tilde{D}(k,a)}{D(a)}=1-\frac{\Omega_b}{\Omega_m}\Big[1-\frac{a^*}{a}
\frac{\tilde{\Delta}_k}{\Delta_k^*}\Big] \,.
\end{equation}
Thus for a given $k$-mode with $k>\tilde{k}_J^*$, the
plasma-suppressed matter power spectrum relative to that of the
conventional approach reads, for $k > \tilde{k}_J(a)>
\tilde{k}_J^*$,
\begin{eqnarray}
\Big|\frac{\tilde{D}(k,a)}{D(a)}\Big|^2
        \simeq \Big(\frac{\Omega_{dm}}{\Omega_{m}}\Big)^2
        &\Big\{&1+2\Big(\frac{\Omega_{b}}{\Omega_{dm}}
         \frac{a^*}{a}\Big)e^{-0.8\omega_r\eta}\cos(2\omega_r\eta)\nonumber
         \\
        &+&\Big(\frac{\Omega_{b}}{\Omega_{dm}}
        \frac{{a^*}}{a}\Big)^2e^{-1.6\omega_{r}\eta}\Big\}.\label{suppression1}
\end{eqnarray}
On the other hand we have, for $\tilde{k}_J^*< k <
\tilde{k}_J(a)$,
\begin{eqnarray}
\Big|\frac{\tilde{D}(k,a)}{D(a)}\Big|^2
        &\simeq& \Big(\frac{\Omega_{dm}}{\Omega_{m}}\Big)^2
        \Big\{1+2\Big(\frac{\Omega_{b}}{\Omega_{dm}}
         \frac{a^*}{a}\Big)\Big[1+\frac{a}{a^*}-\Big(\frac{k}{\tilde{k}_J^*}\Big)^{2}\Big]\nonumber\\
        &+&\Big(\frac{\Omega_{b}}{\Omega_{dm}}
        \frac{{a^*}}{a}\Big)^2\Big[1+\frac{a}{a^*}-\Big(\frac{k}{\tilde{k}_J^*}\Big)^{2}\Big]^2\Big\}.\label{suppression2}
\end{eqnarray}

\begin{figure}[htb]
\includegraphics[width=8.5cm]{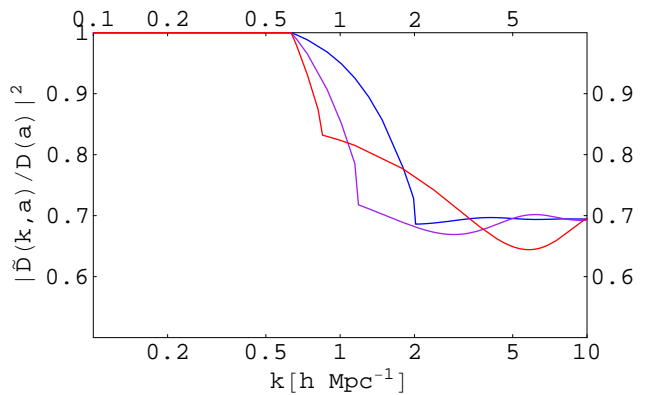}
\caption{\label{Onion_Supp} Suppression of the matter power
spectrum. $a^*=0.1$ and $T^*_b=5\times 10^6{\rm K}$. The red,
purple and blue curves represent the suppressions at $a=1/6, 1/3$
and 1, respectively. The three regimes of $k$ can be clearly
recognized.}
\end{figure}

\begin{figure}[htb]
\includegraphics[width=8.5cm]{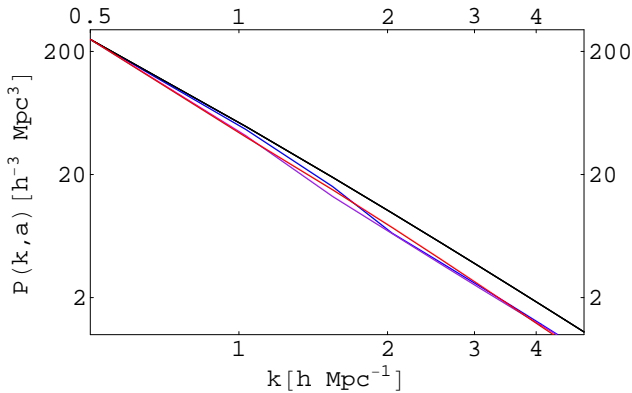}
\caption{\label{Onion} Matter power spectrum with and without the
plasma effect. The conventions are the same as that in
Fig.\ref{Onion_Supp}.}
\end{figure}

To appreciate what this implies in cosmology, we look for the
physical scales relevant to the plasma suppression effect. Let us
assume that the reionization occurred at $1+z^*=10$, or $
a^*=0.1$, with the initial plasma temperature $T_e^*\simeq
T_b^*\sim 5\times 10^6 {\rm K}$. This corresponds to an
ion-acoustic wave velocity of $\tilde{c}_{s}^*\sim 7 \times
10^{-4}c$ and $\tilde{k}_J^*\sim 0.63 h{\rm Mpc}^{-1}$, and
therefore the Jean's length, $\tilde{\lambda}_J^*=
2\pi/\tilde{k}_J^*\sim 1.0 h^{-1}{\rm Mpc}$. The Jean's scale at
present would be $\tilde{k}_J(a=1)\sim 2.0 h{\rm Mpc}^{-1}$, or
$\tilde{\lambda}_J(a=1)\sim 3.1 h^{-1}{\rm Mpc}$. Figure
\ref{Onion_Supp} plots the suppression factor,
$|\tilde{D}(k,a)/D(a)|^2$, at different times: $a=1/6, 1/3$ and 1,
based on the $\Lambda$CDM model with $\Omega_{\Lambda}=0.7,
\Omega_m=0.3, \Omega_{dm}=0.25, \Omega_b=0.05, a^*=0.1, a=1$ and
$T_b^*=5\times 10^6 {\rm K}$. The kinks of each curve are
associated with the transitions at $\tilde{k}_J^*$ and
$\tilde{k}_J(a_c)$, respectively. These kinks are the artifact of
our approximation, which should be smeared in reality. The three
regimes of $k$, namely, $k<\tilde{k}_J^*$,
$\tilde{k}_J(a=1)>k>\tilde{k}_J^*$, and
$k>\tilde{k}_J(a=1)>\tilde{k}_J^*$, can be readily recognized in
this plot. For a mode at $k=\tilde{k}_J(a=1)=2.0h{\rm Mpc}^{-1}$,
for example, it would continue to oscillate until now. Its
oscillation `frequency' is $\omega_r\sim 0.76$. Thus
$2\omega_r\eta\sim 3.3=0.52(2\pi)$, i.e., it would have oscillated
for roughly half a cycle. By then its amplitude would be
Landau-damped by more than one e-fold. In the last regime, the
suppression reaches an asymptotic value of $\sim 70\%$.
Fig.\ref{Onion} compares the conventional and the
plasma-suppressed matter power spectra. We have replaced the shape
parameter $\Gamma=\Omega_mh$ in the BBKS formula\cite{BBKS} for
the transfer function by the Eisenstein-Hu\cite{Eisenstein}
rescaled factor. Thus the baryon effects prior to the reionization
epoch has been included in the transfer function, which is common
to both cases.

By including the long-range residual electrostatic potential, we
have shown that the growth of the baryon density contrast during
the reionization epoch should be suppressed by the plasma
collective effect in the form of ion-acoustic oscillations. Though
subject to Landau damping, these oscillations happen to have both
the period and the damping time comparable to the Hubble time.
Thus there should exist certain imprints of these oscillations in
the large scale structure formation history. In the linear regime
of density contrast growth and the asymptotic limit where the
oscillations damp away, the amount of suppression of the matter
power spectrum due to the plasma effect reaches $\sim 30\%$, which
is sizable. In principle the plasma suppression is also effective
in the nonlinear regime. It would be interesting to expand our
treatment to the nonlinear regime of growth at late times ($a\to
1$) and track its impact.

One major effort in modern cosmology is to understand the cosmic
composition through the determination of a set of cosmic
parameters. In addition to the CMB fluctuations, large scale
structure formation can help further constrain these parameters.
In this regard, our predicted plasma suppression of the power
spectrum in the Lyman-$\alpha$ forest region may be relevant. We
note that the observed power spectrum in the inter-galactic medium
(IGM) region indeed appears to be systematically lower than that
predicted by the conventional theory\cite{Tegmark2}. It would be
very interesting to confirm this. Aside from this issue, the
evolution of the plasma-suppressed power spectrum should provide a
unique window to reveal the detail history and dynamics of
reionization.

\acknowledgements We thank T. Abel, R. Blandford, W. Hu, Ue-Li
Peng, J. Silk and R. Wechsler for helpful discussions. This work
is supported in part by US Department of Energy under Contract
No.\ DE-AC03-76SF00515.

\end{document}